\documentclass[prl,twocolumn,
                                groupedaddress,
                                showpacs,
                                ]{revtex4-1}
\usepackage{graphicx}
\usepackage{amsmath}
\usepackage{amsfonts}
\usepackage{amssymb}
\usepackage{hyperref}
\usepackage{color}

\newcommand{\beq}{\begin{equation}}
\newcommand{\eeq}{\end{equation}}

\usepackage{placeins} 
\usepackage{bm}         

\begin{document}

\title{Dissipation-assisted prethermalization in long-range interacting atomic ensembles}

\author{Stefan Sch\"utz} 
\affiliation{Theoretische Physik, Universit\"at des Saarlandes, D-66123 Saarbr\"ucken, Germany} 

\author{Simon B. J\"ager} 
\affiliation{Theoretische Physik, Universit\"at des Saarlandes, D-66123 Saarbr\"ucken, Germany}

\author{Giovanna Morigi} 
\affiliation{Theoretische Physik, Universit\"at des Saarlandes, D-66123 Saarbr\"ucken, Germany} 

\date{\today}

\begin{abstract}	
We theoretically characterize the semiclassical dynamics of an ensemble of atoms after a sudden quench across a driven-dissipative second-order phase transition. The atoms are driven by a laser and interact via conservative and dissipative long-range forces mediated by the photons of a single-mode cavity. These forces can cool the motion and, above a threshold value of the laser intensity, induce spatial ordering. We show that the relaxation dynamics following the quench exhibits a long prethermalizing behaviour which is first dominated by coherent long-range forces, and then by their interplay with dissipation. Remarkably, dissipation-assisted prethermalization is orders of magnitude longer than prethermalization due to the coherent dynamics. We show that it is associated with the creation of momentum-position correlations, which remain nonzero for even longer times than mean-field predicts. This implies that cavity cooling of an atomic ensemble into the selforganized phase can require longer time scales than the typical experimental duration. In general, these results demonstrate that noise and dissipation can substantially slow down the onset of thermalization in long-range interacting many-body systems.
\end{abstract}

\pacs{37.30.+i, 42.65.Sf, 05.65.+b, 05.70.Ln}
% 37.30.+i 	Atoms, molecules, and ions in cavities 
% 42.65-Sf 	Dynamics of nonlinear optical systems; optical instabilities, optical chaos and complexity, and optical spatio-temporal dynamics 
% 05.65.+b	Self-organized systems 
% 05.70.Ln 	 Nonequilibrium and irreversible thermodynamics 

\maketitle

The quest for a systematic understanding of non-equilibrium phenomena is an open problem in theoretical physics for its importance in the description of dynamics from the microscopic up to astrophysical scales \cite{Zia,Levin,Campa:2009}. Aspects of these dynamics are studied in the relaxation of systems undergoing temporal changes (quenches) of the control field across a critical point \cite{Calabrese,Cugliandolo,Polkovnikov:2011}. Quenches across a non-equilibrium phase transition provide further insight into the interplay between noise and external drives on criticality and thermalization \cite{Hohenberg,Diehl:2015}. In this context photonic systems play a prominent role, thanks to their versatility \cite{Tomadin:2010,Walther:2012,Keeling,Carusotto:2013,Rabl:2014,Peano:2015,Ritsch:2013}. 

\begin{figure}[htbp] %[b]
	\begin{center}
		\includegraphics[width=0.485\textwidth]{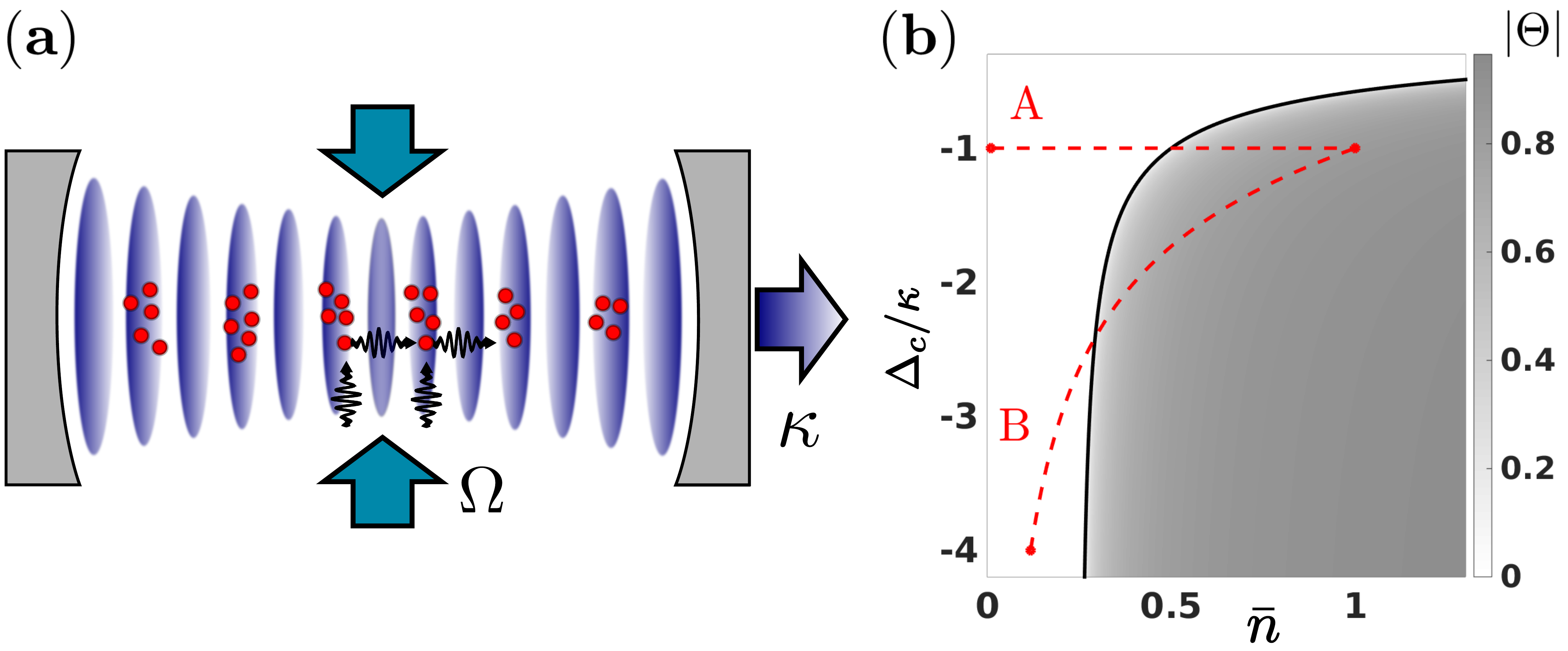}
		\caption{(Color online) (a) Atoms interact with the standing-wave mode of a cavity and are transversally driven by a laser. The laser amplitude ($\Omega$) and/or frequency ($\Delta_c$) are suddenly quenched across the threshold, above which the atoms organize in regular spatial patterns at steady state. The coherent scattering amplitude per atom, $S$, is tuned by the laser, $S\propto \Omega$, the resonator dissipates photons at rate $\kappa$. (b) Phase diagram of the second-order self-organization transition as a function of $\bar{n}$ (proportional to $S^2$) and $\Delta_c/\kappa$ (that determines the asymptotic temperature).  The black line separates the homogeneous phase (with order parameter $\Theta=0$) from the self-organized one (with $\Theta\to \pm 1$). The red dashed lines A and B illustrate the initial and final values of the sudden quenches we analyse.}
		\label{fig:sys}
	\end{center}
\end{figure}

Polarizable particles in a high-finesse cavity, like in the setup illustrated in Fig. \ref{fig:sys}(a), offer a unique system to study relaxation in long-range interacting systems. Here, multiple photon scattering mediates particle-particle interactions whose range scales with the system size in a single-mode cavity \cite{Ritsch:2013,Schuetz:2014,Piazza:2015,Tesio:2014}. In this limit, atomic ensembles in cavities are expected to share several features with other long-range interacting systems such as gravitational clusters and non-neutral plasmas in two or more dimensions \cite{Campa:2009,Cataliotti:2012,Schuetz:2014}. The equilibrium thermodynamics of these systems can exhibit ensemble inequivalence \cite{Campa:2009,Latella:2015}, while quasi-stationary states (QSS) typically characterise the out-of-equilibrium dynamics \cite{Antoni:1995,Campa:2009,Joyce:2010,Filho:2014}. QSS are metastable states in which the system is expected to remain trapped in the thermodynamic limit, they are Vlasov-stable solutions and thus depend on the initial state. So far, however, evidence of QSS has been elusive. It has been conjectured that noise and dissipation can set a time scale that limits the QSS lifetime \cite{Gupta:2010,Chavanis:2011,Bouchet:2012,Chavanis:2014}, and possibly gives rise to dynamical phase transitions \cite{Chavanis:2011}. In Ref. \cite{Joyce:2015} it was shown that, in presence of dissipation due to viscous damping or local inelastic collisions, the relaxation dynamics of long-range interacting systems can be cast in terms of so-called scaling QSS, which are solutions of the kinetic mean-field equation and asymptotically tend to a unique QSS \cite{Joyce:2015}. Accordingly, one would expect to observe QSS in cavity systems \cite{Cataliotti:2012}. In Ref. \cite{Schuetz:2014}, however, we found no evidence of the typical superlinear dependence on $N$ of the QSS time scale \cite{Campa:2009}, which we attributed to the effect of noise and dissipative processes.  Nonetheless, the dissipative dynamics is here due to retardation effects in the coupling between the atoms and a global variable, the cavity field, and can also establish long-range correlations \cite{Asboth:2004,Schuetz:2013} whose influence on the relaxation dynamics is still unexplored.

In this work we characterise the interplay between dissipative and conservative long-range forces in the semiclassical dynamics of $N$ polarizable particles (atoms) confined within a high-finesse single-mode cavity and transversally driven by a laser \cite{Schuetz:2014,Schuetz:2015,Domokos:2002,Black:2003} (see Fig. \ref{fig:sys}(a)). The particles motion is along the cavity axis ($x$-axis), and the dynamics results from their opto-mechanical coupling with the cavity mode at wave number $k$ and spatial mode function $\cos(kx)$. We focus on the regime where the laser frequency $\omega_L$ is smaller than the cavity frequency $\omega_c$, such that $\Delta_c=\omega_L-\omega_c<0$. Here, the dynamics is characterised by a thermal stationary state, which can exhibit a second-order driven-dissipative phase transition (spatial selforganization) as a function of the laser intensity and of $\Delta_c$ \cite{Schuetz:2015}. This transition is due to the interplay between the dispersive and the dissipative forces: The dispersive forces tend to order the atoms in gratings for which the order parameter $\Theta=\sum_{j=1}^N\cos(kx_j)/N\to \pm 1$, with $x_j$ the particles positions, and the intracavity photon number is maximum. The dissipative forces, instead, are due to retardation effects in the dynamics of atoms and field: For $\Delta_c<0$ they cool the atoms  into a thermal state whose effective temperature $T_{\rm eff}$ is determined by $\Delta_c$ and by the cavity loss rate $\kappa$:  $k_BT_{\rm eff}=\hbar(\Delta_c^2 + \kappa^2)/(- 4 \Delta_c)$, with $k_B$ Boltzmann constant  \cite{Schuetz:2015,Niedenzu:2012,Domokos:2002,Schuetz:2013,Horak:1997,Vuletic:2000,Polarizable}. $T_{\rm eff}$ determines the threshold $S_c$ of the coherent laser scattering amplitude $S$ per atom at which spatial self-organization occurs, such that $\sqrt{N}S_c =2k_BT_{\rm eff}/\hbar$ \cite{Schuetz:2015,Asboth:2005,Niedenzu:2012} and separates the regime where the spatial distribution is uniform and $\Theta\simeq 0$ from the symmetry broken phase in which the atoms form Bragg gratings, as shown in Fig. \ref{fig:sys}(b). 

We analyse the semiclassical dynamics of the atoms after a quench across the transition using a Fokker-Planck equation (FPE) for the phase space distribution $f(x_1,\ldots,x_N;p_1,\ldots,p_N;t)$ at time $t$ and as a function of the atoms positions $x_j$ and the momenta $p_j$. The FPE is valid when the cavity linewidth $\kappa$ exceeds the recoil frequency $\omega_r=\hbar k^2/(2m)$ and the width of the momentum distribution $\Delta p$ is larger than the photon linear momentum $\hbar k$ \cite{Schuetz:2013}. It reads \cite{Schuetz:2015,footnote}
\begin{align}
\partial_t f = \{H, f  \} + \bar{n}\mathcal{L}_\beta  f +{\rm O}(U_0),\,  \label{eq:FPE}
\end{align}
where Hamiltonian $H = \sum_{j=1}^N p_j^2/(2m) + \hbar \Delta_c \bar{n} N \Theta^2$ determines the coherent dynamics and is a realization of the anisotropic Hamiltonian Mean Field model (HMF) \cite{Antoni:1995,Jain:2007,Schuetz:2014}. %For $\Delta_c<0$ the potential term tends to localize the atoms within ordered patterns that maximize $\Theta$, and thus the depth of the resulting collective potential. 
The dimensionless parameter $\bar n=N S^2/ (\kappa^2+ \Delta_c^2)$ scales the depth of the conservative potential. It also scales the dissipator $\mathcal{L}_\beta$, describing the effective long-ranged friction and diffusion \cite{Schuetz:2013,Schuetz:2015}:
\begin{align}
\mathcal{L}_\beta f = \sum_{i}^{N} \frac{ \Gamma}{N} \sum_{j}^{N} \sin(kx_i) \partial_{p_i} \sin(kx_j) \left( p_j + \frac{m}{\beta} \partial_{p_j} \right) f\,, 
\label{eq:dissip}
\end{align}
with $\Gamma= 2 \omega_r \hbar \kappa \beta$ and $\beta=(k_BT_{\rm eff})^{-1}$. For $\Delta_c<0$ the incoherent dynamics drives the system into the stationary state $f_S(\beta,\bar n) = f_0 \exp(- \beta H)$, where $f_0$ warrants normalization. This state is well defined in the thermodynamic limit we choose, according to which as $N$ is increased, the quantity $NS^2$ (and thus $\bar n$) is kept constant. This choice warrants that the Hamiltonian satisfies Kac's scaling \cite{Campa:2009}.

The relaxation dynamics following a sudden quench at $t=0$ is numerically evaluated by means of stochastic differential equations (SDE). Averages are taken over several trajectories, sampling the dynamics of $N$ atoms according to the given initial distribution \cite{Domokos:2001,Schuetz:2013}. Before the quench is performed ($t<0$), we assume that the system has reached the equilibrium solution $f_S(\beta,\bar n_i)$ of the FPE at a given value of $\bar n=\bar n_i$ and $\Delta_c$. At $t=0$ the value of $\bar n$ is quenched from $\bar n_i<\bar n_c$, deep in the disordered phase, to $\bar n_f> \bar n_c$, well inside the ordered phase. This corresponds to the horizontal path A of Fig. \ref{fig:sys}(b), keeping $\Delta_c$, and hence the asymptotic temperature, constant. 
We evolve the initial state setting $\bar n=\bar n_f$ in Eq. \eqref{eq:FPE}. In what follows we focus on quenches from the disordered to the ordered phase along path A, nevertheless the essential features of the dynamics we will report on characterize also the quenches in the opposite direction as well as along paths of type B, which connects points with different asymptotic temperatures (see Supplemental Material, SM, \cite{supp}).

\begin{figure} [htbp]%[htbp]
	\begin{center}
		\includegraphics[width=0.4\textwidth]{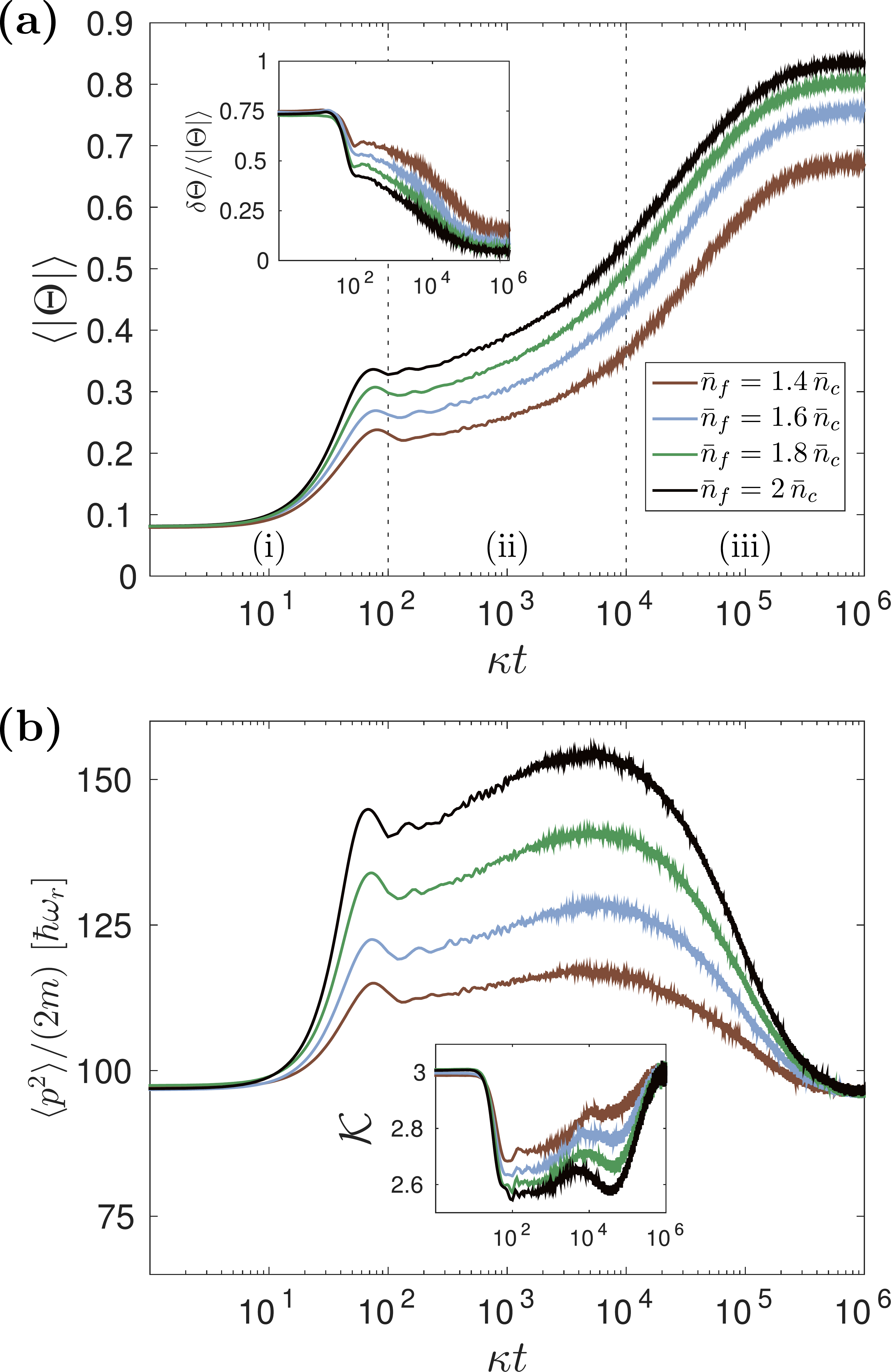}
		\caption{
			(Color online) Numerical simulation of the dynamics following a sudden quench along path A using the SDE \cite{Schuetz:2015}. At $t=0$ the atoms are in the stationary state of Eq. \eqref{eq:FPE} for $\bar{n}_i=0.01  \bar{n}_c$ with $\Delta_c=-\kappa$ and $\bar n$ is quenched to the value $\bar n_f>\bar n_c$ (see legenda in (a)). (a) The modulus of the order parameter $\langle |\Theta| \rangle$ and (b) the single-particle kinetic energy $\langle p^2/(2m)\rangle$ (in units of $\hbar\omega_r$) as a function of time (in units of $\kappa^{-1}$) for $N=50$. The corresponding insets display the time evolution of the relative localization $\delta\Theta/\langle |\Theta| \rangle$, where $\delta\Theta=\sqrt{\langle\Theta^2\rangle-\langle|\Theta|\rangle^2}$, and of the Kurtosis ${\mathcal K}$. The initial values $\langle |\Theta| \rangle_{t=0} \simeq 1/\sqrt{\pi N} \approx 0.08$ in (a) are due to finite $N$ \cite{Schuetz:2015}. Here, $\kappa \approx 390 \,\omega_r$ and $N |U_0| = 0.05 \, \kappa$.  The three relaxation stages are indicated by the labels (i),(ii),(iii).}
		\label{fig:quenchn}
	\end{center}
\end{figure}

The time evolution of the modulus of the order parameter $\langle|\Theta |\rangle$ is displayed in Fig. \ref{fig:quenchn}(a) for different values of $\bar n_f$: $\langle|\Theta |\rangle$ tends towards an asymptotic value, that is closer to unity the larger is $\bar n_f$. Before reaching the steady state the dynamics go through different stages, which we classify as: (i) a fast relaxation towards an intermediate value of the magnetization with time scale $ t \lesssim 10^2\kappa^{-1}$; this time scale decreases with $\bar n_f$. (ii) A transient regime where $\langle|\Theta |\rangle$ seems to grow logarithmically with time. (iii) Finally, the dissipation becomes dominant and brings the system to the asymptotic value, which is exponentially approached over time scales of the order of $10^6\kappa^{-1}$. These time scales are illustrated in Fig. \ref{fig:quenchn}(a) and here reported for $N=50$ particles but generally depend on $N$, as we discuss later on. 

%\begin{widetext}
%\end{widetext}
%\FloatBarrier

\begin{figure}[htbp]
	\begin{center}
		\includegraphics[width=0.4\textwidth]{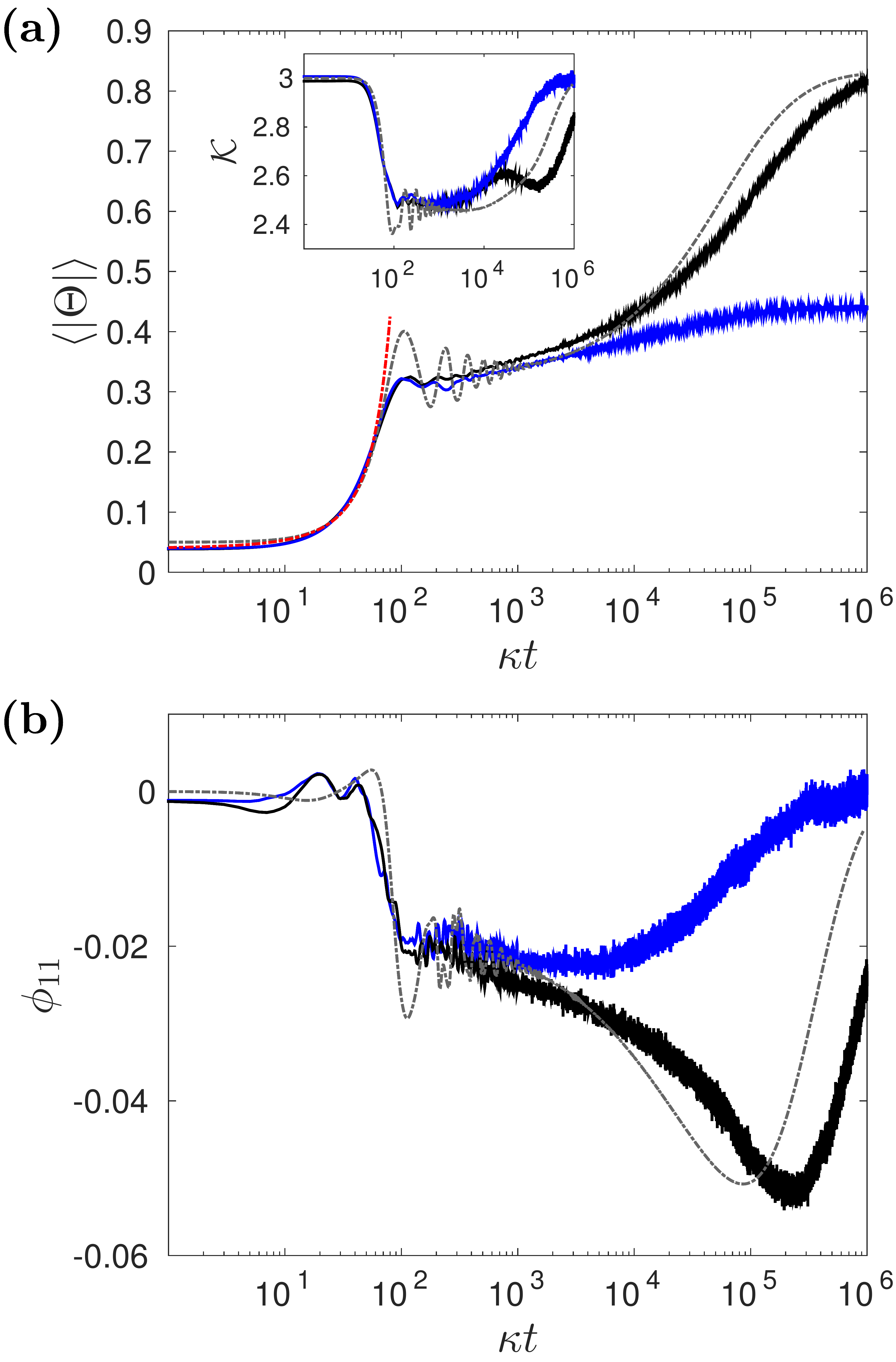}
		\caption{(Color online) 
			Dynamics following a sudden quench along path A with $\bar{n}_f = 2 \bar{n}_c$ and $N=200$. At $t=0$ the atoms are in the stationary state of Eq. \eqref{eq:FPE} for $\bar{n}_i=0.01  \bar{n}_c$ and $\Delta_c=-\kappa$. 
			Subplot (a) compares the evolution of $\langle |\Theta| \rangle$ and ${\mathcal K}$ (inset) obtained by integrating Eq. \eqref{eq:FPE} (black line), with the one found after setting $\Gamma=U_0=0$ (blue line). Onset: The red line is the fit obtained by a stability analysis of the homogeneous Vlasov solution, the dashed-dotted line by a mean-field model (see SM \cite{supp}).
			(b) Time evolution of the QSS observable $\phi_{11}$, Eq. \eqref{phi11}, corresponding to the curves in (a).}
		\label{fig:QSS}
	\end{center}
\end{figure}

We first observe that, being $\Delta_c$ negative, the growth of $\langle|\Theta |\rangle$, Fig. \ref{fig:quenchn}(a), corresponds to a monotonic decrease of the potential energy, $V= \hbar \Delta_c \bar{n} N \Theta^2$. In the fast relaxation  stage (i), this decrease is well-fitted by an exponential, and is associated with a corresponding decrease of the relative fluctuations (see inset), indicating that the cavity field exponentially grows and creates a mechanical potential, which increasingly localizes the atoms at its minima. The exponential potential depth growth is due to this nonlinearity: the more the atoms become localized in the Bragg grating the larger is the scattering amplitude, and thus the potential depth. The increasing localization correspondingly augments the kinetic energy, as visible in Fig. \ref{fig:quenchn}(b). In this regime, thus, the total energy is conserved, the dynamics is coherent and consists in a transfer of energy from spatial into momentum fluctuations. Correspondingly, the single-particle momentum distribution becomes increasingly non-thermal, as visible by inspecting the time-evolution of the Kurtosis, ${\mathcal K}=\langle p^4 \rangle / \langle p^2 \rangle^2$, shown in the inset of Fig. \ref{fig:quenchn}(b): ${\mathcal K}$ exponentially deviates from the value of the initial Gaussian ("thermal") state, for which ${\mathcal K}_{\rm gauss}=3$. We have verified that this dynamics is  well described by a Vlasov equation for the single-particle distribution $f_1(x,p;t)$, which we derive assuming $f(x_1,\ldots,x_N;p_1,\ldots,p_N;t)=\prod_{j=1}^Nf_1(x_j,p_j;t)$, integrating out the $N-1$ variables from Eq. \eqref{eq:FPE} for the initial uniform distribution and taking the thermodynamic limit (see SM \cite{supp} and \cite{Simon}). Figure \ref{fig:QSS}(a) compares the result of the FPE with the predictions of the Vlasov equation (red curve), showing an excellent agreement in the fast relaxation regime. Numerical and analytical results  show that the time scale of this dynamics depends on $N$ only through the parameter $\bar n$ (and is thus constant when Kac's scaling applies), see also SM \cite{supp}.

After this fast relaxation, the growth in the order parameter and in the kinetic energy seems logarithmic in time. This transient regime (ii) is of Hamiltonian origin: It exhibits damped oscillations, which can be understood as oscillations of the atoms at the minima. Energy is periodically transferred from the kinetic to the potential energy. Since the potential energy depends on a global variable, energy is exchanged between the particles by means of elastic collisions,
hence damping the oscillations.  Correspondingly, the Kurtosis starts to increase towards the Gaussian value, showing that the sample starts to equilibrate. In order to verify this hypothesis, in Fig. \ref{fig:QSS}(a) we compare the predictions of the full simulation (black curves) for order parameter and Kurtosis with the ones obtained after setting $\Gamma=0$ in Eq. \eqref{eq:FPE} (blue curves): in the transient regime the curves nearly overlap for $t\lesssim 10^{4}\kappa^{-1}$. Noise and dissipation, however, lead to a discrepancy between the predictions of the Hamiltonian and of the full FPE. This discrepancy becomes increasingly evident at longer time scales: When the dynamics is solely Hamiltonian, in fact, the Kurtosis increases monotonically towards the Gaussian value. Due to the analogy with the Hamiltonian dynamics, some of the features of the transient regime are reminiscent of the HMF, where for a similar quench a violent relaxation is observed, then followed by prethermalization in a QSS \cite{Antoni:1995,Jain:2007}. In our case, for $\Gamma \neq 0$,  as in Ref. \cite{Schuetz:2014}, we do not find evidence of a superlinear scaling with $N$ of the QSS lifetime. The QSS lifetime, in fact, is limited by the dissipative effects, which have the same physical origin as the long-range conservative forces and whose characteristic time scale is linear in $N$ (see SM, \cite{supp}, and Ref. \cite{Simon}). Note that at the end of this stage the atoms are localised, but their temperature is hotter than $T_{\rm eff}$.

\begin{figure}[htbp]
	\begin{center}
	\includegraphics[width=0.44\textwidth]{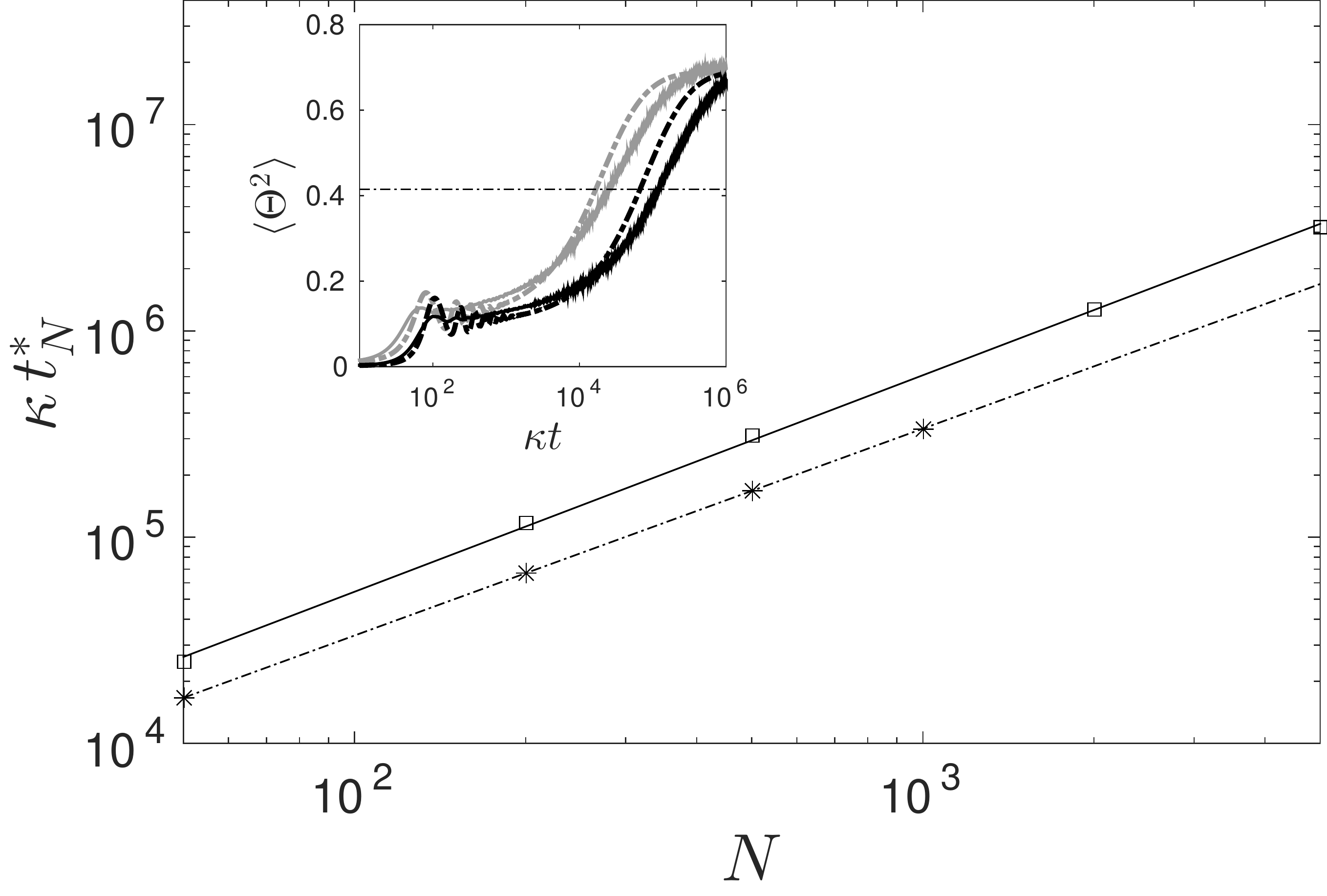}
	\caption{
Relaxation after a sudden quench along path A for the parameters of Fig \ref{fig:QSS} ($\bar{n}_f = 2 \bar{n}_c$) and for different $N$. Inset: Time evolution of $\langle \Theta^2\rangle$ for $N=50$ (gray) and $N=200$ (black). The dashed lines are the prediction of the MF model (see SM, \cite{supp}). The horizontal dashed line indicates where $\langle \Theta^2 \rangle$ has reached $60 \%$ of its stationary value and identifies the corresponding time $t^*_N$. Onset: time $t^*_N$ (in units of $\kappa^{-1}$) as a function of $N$ for the full FPE (squares) and MF (stars). The lines are the corresponding linear fits in the log-log plot. 
%(color online) 
%Onset: Time evolution of $\langle \Theta^2\rangle$ for $N=50$ (gray) and $N=200$ (black) after a sudden quench along path A. The dashed line reports the prediction of the mean-field model (see SM, \cite{supp}). The parameters are the same as in Fig. \ref{fig:QSS} ($\bar{n}_f = 2\, \bar{n}_c$). The horizontal dashed line indicates when the order parameter has reached $60 \%$ of its stationary value. The time $t^*_N$ is at the point where this line crosses the curves. Inset: time $t^*_N$ as a function of $N$ for the full FPE (squared) and mean-field (stars). The lines are the respective linear fits in the log-log plot. The time axis is in units of $\kappa^{-1}$.
}
	\label{fig:scaling}
    \end{center}
\end{figure}

In stage (iii), when the effect of dissipation becomes relevant, the atoms are cooled and further localised at the minima. The Kurtosis, however, further decreases till reaching a minimum, before increasing again towards the Gaussian value. We first compare this behaviour with the predictions of a mean-field (MF) model, which we extract from Eq. \eqref{eq:FPE}  by means of the factorization ansatz, see SM \cite{supp}. The grey lines in Fig. \ref{fig:QSS}(a) and its inset show the MF predictions as a function of time and indicate that, even though MF reproduces qualitatively the dynamical features, it fails to give the correct time scale by at least one order of magnitude. Further insight is provided by the observable for QSS \cite{Joyce:2015}, which we here define as: 
\begin{equation}
\label{phi11}
\phi_{11} = \frac{\langle |\sin(kx) p| \rangle }{ \langle |\sin(kx)| \rangle \langle |p| \rangle } -1\,.
\end{equation}
When $\phi_{11}\neq 0$, the distribution is not factorizable into a kinetic and a potential term. Figure \ref{fig:QSS}(b) displays the time evolution of $\phi_{11}$ for the Hamiltonian, mean-field, and full dynamics. In stages (i) and (ii)  the three models predict approximately the same behaviour. Instead, in stage (iii), $\phi_{11}$ evolves differently: For both MF and full FPE it exhibits a minimum, however reached at different times, which seems to possess the features of a scaling
QSS, namely, a sequence of QSS with identical correlations \cite{Joyce:2015}. Its nature could be understood in terms of the onset of collective oscillations which are (almost) decoupled from noise and dissipation. Analogous behaviours have been reported for the case of atomic arrays in a cavity \cite{Asboth:2004,Mishina:2014}. Since the trajectories of $\phi_{11}$ are different for the three types of simulations, the corresponding QSS are expected to not be the same. In particular, the discrepancy between full FPE and MF in stage (iii) remains of the same order when scaling up the system, while instead Hamiltonian prethermalization tends towards the corresponding mean-field prediction. Figure  \ref{fig:scaling} displays the relaxation time scales for the MF and the full FPE: the two curves suggest a linear increase with $N$ for both cases, nevertheless they run parallel thus showing that the discrepancy is a scalable effect. We deduce that this discrepancy is due to the momentum-position correlations due to noise, which are otherwise discarded in the MF treatment. 

This prethermalization is not related to the critical slowing down observed in Ref. \cite{Keeling:2012}, but is due to the creation of correlations between momentum and position, and is reminiscent of kinetic-stop dynamics \cite{Olmos:2012}.  It implies that cavity-cooling of a large sample of atoms into the self-organized phase, corresponding to a sudden quench along path B, can be very slow and thus inefficient (see also Ref. \cite{Niedenzu:2012}).  
Our analysis sets the stage for the development of a kinetic equation that is valid in the full quantum regime \cite{Tureci,Piazza:2014,Nagy,ETH,Hamburg}.

%\acknowledgements
\begin{acknowledgements}
We acknowledge discussions with Ralf Betzholz, Cesare Nardini, and Stefano Ruffo, and support by the German Research Foundation (DFG,
DACH project "Quantum crystals of matter and light")  and by the German Ministry of Education and Research (BMBF "Q.com"). 
\end{acknowledgements}

%\end{document} % %end main.tex

% % %
\newpage
\setcounter{equation}{0}
\setcounter{figure}{0}
% % %

% % % supp.tex \begin{document}-\end{document}
\renewcommand*{\citenumfont}[1]{S#1}
\renewcommand*{\bibnumfmt}[1]{[S#1]}
\renewcommand{\thesection}{S.\arabic{section}}
\renewcommand{\thesubsection}{\thesection.\arabic{subsection}}
\makeatletter %% With ams
\def\tagform@#1{\maketag@@@{(S\ignorespaces#1\unskip\@@italiccorr)}}
\makeatother
\makeatletter
\makeatletter \renewcommand{\fnum@figure}
{\figurename~S\thefigure}
\makeatother

\onecolumngrid
\begin{center}
	\textbf{\large Supplemental Material for\\
		Dissipation-assisted prethermalization in long-range interacting atomic ensembles}
\end{center}

\subsection{Mean-Field equation and Vlasov limit}
The mean-field equation is derived from Eq. (1) in the main text by assuming $f(x_1,\ldots,x_N;p_1,\ldots,p_N;t)=\prod_{j=1}^Nf_1(x_j,p_j;t)$, and integrating out the $N-1$ variables. It reads 
\begin{align}
\partial_t f_1=&\{H_{\mathrm{MF}}[f_1],f_1\}+\bar{n}\mathcal{L}_{\beta,\mathrm{MF}}f_1 \label{eq:meanFPE}
\end{align}
where the mean-field Hamiltonian $H_{\mathrm{MF}}[f_1]$ is given by
\begin{align*}
H_{\mathrm{MF}}[f_1]=\frac{p^2}{2m}+\frac{2\hbar\Delta_c\bar{n}}{N}\left(\frac{1}{2}\cos(kx)+(N-1)\langle \cos(kx') \rangle_{f_1}\nonumber-\frac{ \hbar k \beta \kappa}{2m \Delta_c}(N-1) \langle p'\sin(kx')\rangle_{f_1}\right)\cos(kx)\,,
\end{align*}
with $\langle \mathcal{A}(x',p') \rangle_{f_1}= \int \limits_0^{\lambda} {\rm d}x' \int \limits_{-\infty}^{\infty} {\rm d} p' \mathcal{A}(x',p') f_1(x',p')$ and $\mathcal{A}$ a phase space function. Furthermore the mean-field dissipator $\mathcal{L}_{\beta,\mathrm{MF}}$ is defined as 
\begin{align*}
\mathcal{L}_{\beta,\mathrm{MF}}f_1=\frac{\Gamma}{N}\sin^2(kx)\partial_p\left(p+\frac{m}{\beta}\partial_p\right)f_1\, .
\end{align*}
The dissipator $\mathcal{L}_{\beta,\mathrm{MF}}$ is responsible for the relaxation of the system to the thermal stationary state with temperature $\beta^{-1}=k_BT_{\rm eff}=\hbar(\Delta_c^2 + \kappa^2)/(- 4 \Delta_c)$ \cite{s1,s2,s3}. %\cite{Schuetz:2014,Schuetz:2015,Jaeger:2016}. 
Since the dissipator decreases with $N^{-1}$ (for increasing $N$), the mean-field predicts a relaxation timescale that extends linearly with $N$. Although the relaxation for the full FPE, Eq. (1) in the main text, is orders of magnitudes slower (see Fig. 4 in the paper), the corresponding growth of the timescale with $N$ is almost indistinguishable from a linear one.\\
In order to make some statements for the short time dynamics we derive the Vlasov equation. This equation is derived from Eq. \eqref{eq:meanFPE} after performing the limit $N\to\infty$ with $NS^2=$const. and reads
\begin{align}
\partial_t f_1 + \frac{p}{m} \partial_{x} f_1 - \partial_x V[f_1] \partial_p f_1 = 0\,,
\label{eq:Vlasov}
\end{align}
where the Vlasov potential $V[f_1]$ is 
\begin{align}
V[f_1] = 2 \hbar \Delta_c \bar{n} \left( \langle \cos(kx') \rangle_{f_1}-\frac{ \hbar k \beta \kappa}{2m \Delta_c} \langle p'\sin(kx')\rangle_{f_1}\right)\cos(kx)\,. \nonumber
\end{align}
The stability analysis of Eq. \eqref{eq:Vlasov} shows that a spatially homogeneous distribution is unstable against small fluctuations $\delta f$ when $\bar{n}_f > \bar{n}_c$. The fluctuations exhibit exponential growth at rate $\gamma$, which %solely depends on the final value of the quench, it 
monotonously increases with $\bar n_f$ and is a solution of the equation
\begin{align}
\big[ 1 - 2\kappa \gamma /(\Delta_c^2 + \kappa^2)\big] F( \gamma)\bar{n}_f /\bar{n}_c=1\,,
\label{eq:viol}
\end{align}
with $F(\gamma)=1- \sqrt{\pi} b \exp(b^2) \text{erfc}(b) $, $b^2 = \hbar \gamma^2 \beta /(4 \omega_r)$ and erfc is the complementary error function. The solution (red line in Fig. 3(a) in the main text for a fixed value of $\bar n_f$) well fits the numerical result in the fast relaxation regime. Thus, this initial behaviour is analogous to the violent relaxation observed in the HMF and has mainly Hamiltonian 
origin. 

\subsection{Quenches along path B}

\begin{figure*}[htbp]
	\begin{center}			
		\includegraphics[width=1\textwidth]{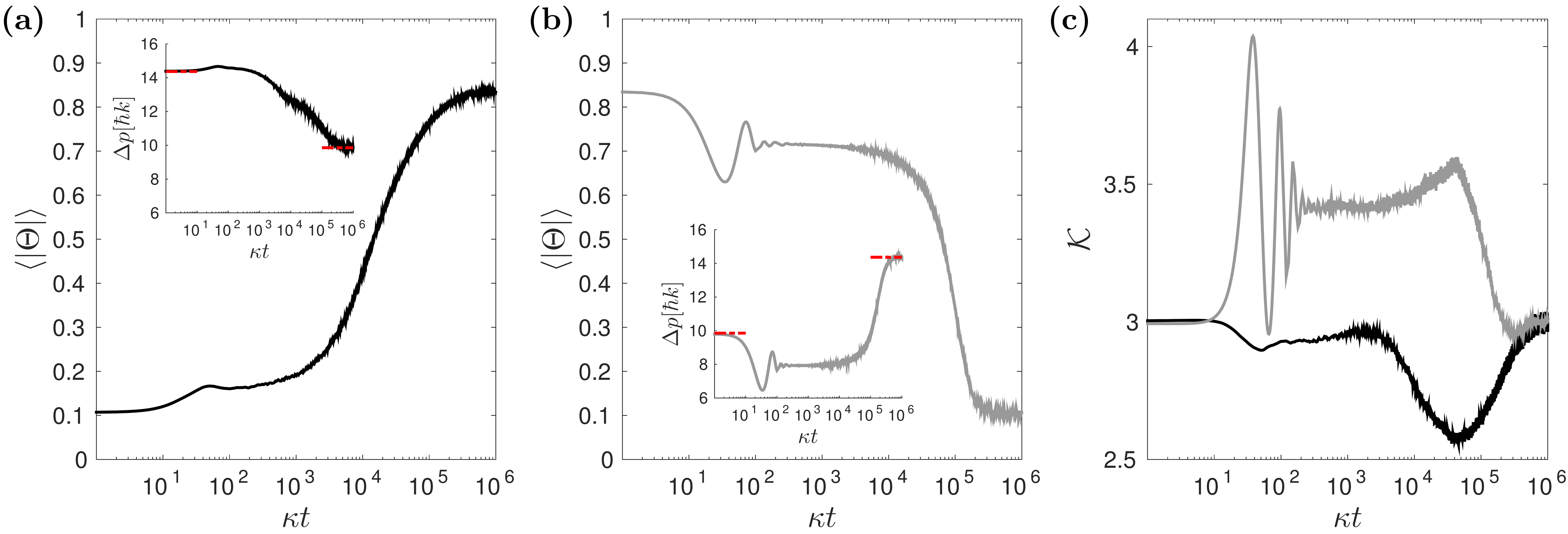}
		\caption{
			Numerical simulation of the dynamics for $N=50$ atoms following a sudden quench along path B, where $\Delta_c$ is varied but the laser intensity is kept fixed. The black line corresponds to the results for the evolution when the value of the detuning is suddenly quenched from $\Delta_c=- 4 \kappa$ with $\bar{n}_i \approx 0.44\, \bar{n}_c$ to $\Delta_c = - \kappa$ with $\bar{n}_f = 2 \,\bar{n}_c$. The grey line displays the case where initial and final points are swapped.  (a) and (b): Time evolution of $\langle |\Theta| \rangle$ and $\Delta p$ (inset) as a function of time (in units of $\kappa^{-1}$). Subplot (c) displays the behaviour of the kurtosis ${\mathcal K}$.
		}
		\label{fig:quenchdc}
	\end{center}
\end{figure*}

Figure S\ref{fig:quenchdc} displays a sudden quench in the detuning $\Delta_c$ while keeping $\Omega$, hence the laser amplitude, constant. This quench corresponds to path B of Fig. 1(b) in the paper and alters both $\bar n$ and $\beta$ in Eq. (1), namely both the asymptotic order and temperature. We consider quenches from the disordered (with $\Delta_c = - 4\kappa$) to the ordered phase (with $\Delta_c = - \kappa$), and vice versa, assuming that the initial state is the asymptotic state of the parameter choice before the quench. Also in this case the three regimes can be identified. Remarkably, for the quench from the ordered to the disordered phase, the system remains for long times trapped in an ordered pattern. The pattern stays stable due to the long-range forces. This transient is further accompanied by a momentum distribution that is narrower than the initial and the asymptotic value, as visible in the inset of subplot (b). On the other hand, the momentum distribution is markedly non-Gaussian, as visible in (c). This behaviour shows that, even if the final value of the parameter $\bar{n}$ is well below threshold and the asymptotic number of intracavity photons $\langle \hat{a}^{\dag}\hat{a}\rangle\approx N\bar{n}\langle\Theta^2\rangle$ is small, yet there is a metastable regime in which the number of intracavity photons is significantly larger, due to the metastable atomic patterns which support superradiant scattering of photons into the resonator until they decay. 

% % % supp.tex \begin{document}-\end{document}


\begin{thebibliography}{99}
	
	\bibitem{Zia}
	%Non-equilibrium statistical mechanics: From a paradigmatic model to biological transport
	T. Chou, K. Mallick, and R. K. P. Zia, Rep. Prog. Phys. {\bf 74}, 116601 (2011).
	
	\bibitem{Levin}
	Y. Levin, R. Pakter, F. B. Rizzato, T. N. Teles, and F. P. C. Benetti, Phys. Rep. {\bf 535}, 1 (2014).
	
	\bibitem{Campa:2009}
	%Statistical mechanics and dynamics of solvable models with long-range interactions
	A. Campa, T. Dauxois, and S. Ruffo, Phys. Rep. {\bf 480}, 57 (2009).
	
	\bibitem{Calabrese}
	%Quantum quenches in extended systems
	P. Calabrese and J. Cardy, J. Stat. Mech. P06008  (2007).
	
	\bibitem{Cugliandolo}
	L. Foini, L. F. Cugliandolo, and A. Gambassi,
	Phys. Rev. B {\bf 84}, 212404 (2011); F. Igl\'oi and H. Rieger,
	Phys. Rev. Lett. {\bf 106}, 035701 (2011). 
	
	\bibitem{Polkovnikov:2011}
	%Colloquium: Nonequilibrium dynamics of closed interacting quantum systems
	A. Polkovnikov, K. Sengupta, A. Silva, and M. Vengalattore,
	Rev. Mod. Phys. {\bf 83}, 863 (2011).
	
	%\bibitem{Halperin:1977}
	%P. C. Hohenberg and B. I. Halperin, Rev. Mod. Phys. {\bf 49}, 435 (1977).
	%
	%\bibitem{Cross:1993}
	%M. C. Cross and P. C. Hohenberg, Rev. Mod. Phys. {\bf 65}, 851 (1993).
	
	\bibitem{Hohenberg}
	P. C. Hohenberg and B. I. Halperin, Rev. Mod. Phys. {\bf 49}, 435 (1977);
	M. C. Cross and P. C. Hohenberg, Rev. Mod. Phys. {\bf 65}, 851 (1993).
	
	\bibitem{Diehl:2015}
	%Keldysh Field Theory for Driven Open Quantum Systems
	L. M. Sieberer, M. Buchhold, and S. Diehl, preprint arXiv:1512.00637 (2015).
	
	\bibitem{Tomadin:2010}
	A. Tomadin and R. Fazio, J. Opt. Soc. Am. B {\bf 27}, A130 (2010).
	
	\bibitem{Walther:2012}
	%Photonic quantum simulators
	A. Aspuru-Guzik and P. Walther, Nature Physics {\bf 8}, 285 (2012).
	
	\bibitem{Keeling}
	%Nonequilibrium Model of Photon Condensation
	P. Kirton and J. Keeling, Phys. Rev. Lett. {\bf 111}, 100404 (2013).
	
	\bibitem{Rabl:2014}
	%Implementation of the Dicke lattice model in hybrid quantum system arrays
	L. J. Zou, D. Marcos, S. Diehl, S. Putz, J. Schmiedmayer, J. Majer, and P. Rabl,
	Phys. Rev. Lett. {\bf 113}, 023603 (2014).
	
	\bibitem{Carusotto:2013}
	%Quantum fluids of light
	I. Carusotto and C. Ciuti, Rev. Mod. Phys. {\bf 85}, 299 (2013).
	
	\bibitem{Peano:2015}
	%Topological Phases of Sound and Light
	V. Peano, C. Brendel, M. Schmidt, and F. Marquardt, Phys. Rev. X {\bf 5}, 031011 (2015).
	
	\bibitem{Ritsch:2013}
	H. Ritsch, P. Domokos, F. Brennecke, and T. Esslinger, Rev. Mod. Phys. {\bf 85}, 553 (2013).
	
	\bibitem{Schuetz:2014}
	S. Sch{\"u}tz and G. Morigi, Phys. Rev. Lett. {\bf 113}, 203002 (2014).
	
	%\bibitem{Strathclyde:2014}
	
	\bibitem{Piazza:2015}
	F. Piazza and H. Ritsch, Phys. Rev. Lett. {\bf 115}, 163601 (2015).
	
	\bibitem{Tesio:2014}
	E. Tesio, G. R. M. Robb, G.-L. Oppo, P. M. Gomes, T. Ackemann, G. Labeyrie, R. Kaiser, and W. J. Firth, Phil. Trans. R. Soc. A
	{\bf 372}, 20140002 (2014).
	
	\bibitem{Cataliotti:2012}
	R. Bachelard, T. Manos, P. de Buyl, F. Staniscia,
	F. S. Cataliotti, G. De Ninno, D. Fanelli, and N. Piovella,
	J. Stat. Mech. (2010) P06009.
	
	\bibitem{Latella:2015}
	%Article: Thermodynamics of Nonadditive Systems
	I. Latella, A. P\'erez-Madrid, A. Campa, L. Casetti, and S. Ruffo, Phys. Rev. Lett. {\bf 114}, 230601 (2015). 
	
	\bibitem{Antoni:1995}
	M. Antoni and S. Ruffo, Phys. Rev. E {\bf 52}, 2361 (1995).
	
	\bibitem{Joyce:2010}
	%Quasistationary States and the Range of Pair Interactions
	A. Gabrielli, M. Joyce, and B. Marcos, Phys. Rev. Lett. {\bf 105}, 210602 (2010).
	
	\bibitem{Filho:2014}
	T. M. Rocha Filho, A. E. Santana, M. A. Amato, and A. Figueiredo, Phys. Rev. E {\bf 90}, 032133 (2014).
	
	\bibitem{Gupta:2010}
	%Slow Relaxation in Long-Range Interacting Systems with Stochastic Dynamics
	S. Gupta and D. Mukamel, Phys. Rev. Lett. {\bf 105}, 040602 (2010). 
	
	\bibitem{Chavanis:2011}
	P.-H. Chavanis, F. Baldovin, and E. Orlandini, Phys. Rev. E {\bf 83}, 040101 (2011).
	
	\bibitem{Bouchet:2012}
	%Kinetic theory of nonequilibrium stochastic long-range systems: Phase transition and bistability
	C. Nardini, S. Gupta, S. Ruffo, T. Dauxois, and F.  Bouchet,
	J. Stat. Mech. (2012) P12010.
	
	\bibitem{Chavanis:2014}
	P. H. Chavanis, Eur. Phys. J. B {\bf 87}, 120 (2014).
	%The Brownian mean field model
	
	\bibitem{Joyce:2015}
	%Quasistationary states in the self-gravitating sheet model
	M. Joyce and T. Worrakitpoonpon
	Phys. Rev. E {\bf 84}, 011139 (2011);
	%Scaling Quasistationary States in Long-Range Systems with Dissipation
	M. Joyce, J. Morand, F. Sicard, and P. Viot,
	Phys. Rev. Lett. {\bf 112}, 070602 (2014); 
	%Attractor non-equilibrium stationary states in perturbed long-range interacting systems 
	M. Joyce, J. Morand, and P. Viot,
	Phys. Rev. E {\bf 93}, 052129 (2016).
	
	\bibitem{Asboth:2004}
	J. K. Asb\'oth, P. Domokos, and H. Ritsch, Phys. Rev. A {\bf 70}, 013414 (2004). 
	
	\bibitem{Schuetz:2013}
	%Cooling of atomic ensembles in optical cavities: semiclassical limit
	S. Sch\"utz, H. Habibian, and G. Morigi, Phys. Rev. A {\bf 88}, 033427 (2013).
	
	\bibitem{Schuetz:2015}
	S. Sch{\"u}tz, S. B. J\"ager, and G. Morigi, Phys. Rev. A  {\bf 92}, 063808 (2015).
	
	
	\bibitem{Domokos:2002}
	P. Domokos and H. Ritsch, Phys. Rev. Lett. {\bf 89}, 253003 (2002). 
	
	\bibitem{Black:2003}
	A. T. Black, H. W. Chan, and V. Vuleti\'{c}, Phys. Rev. Lett. {\bf 91}, 203001 (2003).
	
	\bibitem{Niedenzu:2012}
	W. Niedenzu, T. Grie{\ss}er, and H. Ritsch, Europhys. Lett. {\bf 96}, 43001 (2011).
	
	\bibitem{Horak:1997}
	P. Horak, G. Hechenblaikner, K. M. Gheri, H. Stecher, and H. Ritsch, Phys. Rev. Lett. {\bf 79}, 4974 (1997).
	
	
	\bibitem{Vuletic:2000}
	%%Laser Cooling of Atoms, Ions, or Molecules by Coherent Scattering
	V. Vuleti\'{c} and S. Chu, Phys. Rev. Lett. {\bf 84}, 3787 (2000).
	
	\bibitem{Polarizable}
	%Simultaneous cooling and trapping of atoms by a single cavity-field mode
	A. Vukics and P. Domokos, Phys. Rev. A {\bf 72}, 031401(R) (2005).
	
	\bibitem{Asboth:2005}
	J. K. Asb{\'o}th, P. Domokos, H. Ritsch, and A. Vukics, Phys. Rev. A {\bf 72}, 053417 (2005).
	
	\bibitem{footnote}
	Equation \eqref{eq:FPE} contains also the processes scaling with the dynamical Stark shift $U_0=g^2/\Delta_a$, where $g$ is the cavity vacuum Rabi frequency and $\Delta_a=\omega_L-\omega_a$ the detuning between laser and atomic transition frequency. 
	These processes are here small but accounted for in the numerical simulations \cite{Schuetz:2015}.
	
	\bibitem{Jain:2007}
	K. Jain, F. Bouchet, and D. Mukamel, J. Stat. Mech. P11008 (2007).
	
	\bibitem{Domokos:2001}
	P. Domokos, P. Horak, and H. Ritsch, J. Phys. B {\bf 34}, 187
	(2001).
	
	\bibitem{supp}
	See Supplemental Material for the Quenches along path B in Fig. \ref{fig:sys} and the derivation of the mean-field FPE and the Vlasov equation.
	
	\bibitem{Simon}
	S. B. J\"ager, S. Sch\"utz, and G. Morigi, preprint arXiv:1603.05148 (2016).
	
	\bibitem{Mishina:2014}
	O. S. Mishina, New J. Phys. {\bf 16}, 033021 (2014).
	
	\bibitem{Keeling:2012}
	M. J. Bhaseen, J. Mayoh, B. D. Simons, and J. Keeling,
	Phys. Rev. A  {\bf 85}, 013817 (2012).
	
	\bibitem{Olmos:2012}
	B. Olmos, I. Lesanovsky, and J. P. Garrahan, Phys. Rev. Lett. {\bf 109}, 020403 (2012). 
	
	\bibitem{Nagy}
	%The Open-System Dicke-Model Quantum Phase Transition with a Sub-Ohmic Bath
	D. Nagy and P. Domokos, Phys. Rev. Lett. {\bf 115}, 043601 (2015).
	
	\bibitem{Tureci}
	%Cavity-Mediated Near-Critical Dissipative Dynamics of a Driven Condensate
	M. Kulkarni, B. \"Oztop, and H. E. T\"ureci, Phys. Rev. Lett. {\bf 111}, 220408 (2013).
	
	\bibitem{Piazza:2014}
	%Quantum kinetics of ultracold fermions coupled to an optical resonator
	F. Piazza and P. Strack, Phys. Rev. A {\bf 90}, 043823 (2014). 
	
	
	
	%\bibitem{Baumann:2010}
	%K. Baumann, C. Guerlin, F. Brennecke, and T. Esslinger, Nature (London) {\bf 464}, 1301 (2010).
	%%
	%
	%\bibitem{PNAS:2013}
	%%%Real-time observation of fluctuations at the driven-dissipative Dicke phase transition
	%F. Brennecke, R. Mottl, K. Baumann, R. Landig, T. Donner, and T. Esslinger,
	%PNAS {\bf 110}, 11763 (2013).
	%
	%\bibitem{Wolke:2013}
	%M. Wolke, J. Klinner, H. Ke{\ss}ler, and A. Hemmerich,
	%Science {\bf 337}, 75 (2012).
	%
	%\bibitem{PNAS:2015}
	%J. Klinder, H. Ke{\ss}ler, M. Wolke, L. Mathey, and A. Hemmerich, 
	%PNAS {\bf 112}, 3290 (2015). 
	
	\bibitem{ETH}
	K. Baumann, C. Guerlin, F. Brennecke, and T. Esslinger, Nature (London) {\bf 464}, 1301 (2010);
	F. Brennecke, R. Mottl, K. Baumann, R. Landig, T. Donner, and T. Esslinger, PNAS {\bf 110}, 11763 (2013).
	
	\bibitem{Hamburg}
	M. Wolke, J. Klinner, H. Ke{\ss}ler, and A. Hemmerich, Science {\bf 337}, 75 (2012);
	J. Klinder, H. Ke{\ss}ler, M. Wolke, L. Mathey, and A. Hemmerich, PNAS {\bf 112}, 3290 (2015).
	
	
	
	
\end{thebibliography}

\begin{thebibliography}{99}
	\bibitem{s1} %\bibitem{Schuetz:2014}
	S. Sch{\"u}tz and G. Morigi, Phys. Rev. Lett. {\bf 113}, 203002 (2014).
	\bibitem{s2} %\bibitem{Schuetz:2015}
	S. Sch\"utz, S. B. J\"ager, and G. Morigi, Phys. Rev. A {\bf 92}, 063808 (2015).
	\bibitem{s3} %\bibitem{Jaeger:2016}
	S. B. J\"ager, S. Sch\"utz, and G. Morigi, preprint arXiv:1603.05148 (2016).
\end{thebibliography}
\end{document}